\documentclass[letterpaper, 10 pt, conference]{ieeeconf}  

\IEEEoverridecommandlockouts                              
\usepackage{url}
\usepackage{tabu}
\usepackage{graphicx}

\title{\LARGE \bf
Rural wireless deployments in India
}

\author{Arpit Bajpai\\Technische Universit\"at M\"unchen\\Munich, Germany\\arpit.bajpai@tum.de $^{}$
\thanks{$^{1}$Arpit Bajpai, Technical University of Munich, Germany
        {\tt\small }}%
\thanks{$^{}$
        {\tt\small}}%
}

\begin{document}

\maketitle
\thispagestyle{empty}
\pagestyle{empty}

\begin{abstract}

The Internet provides access to communications, information and opportunity to people. Growth of technology and its benefits should penetrate down to every citizen of a country. Rural Areas without Internet access risk being left behind in the information age. To be competitive, they must have broad and sustainable Internet services. Internet connectivity is especially important in developing countries like India where 67\% of the population hail from rural areas. It is clear that most developing countries have still basic problems such as education, poverty, housing, and health, which are sure priorities for any government, it is important to determine whether an advanced informational infrastructure provides a tool for economic and social development. There are a lot of challenges that need to be considered while providing connectivity to rural areas such as high cost of installations, a dearth of skilled labor, deficiency of reliable power sources, local cooperation and support etc. In this paper, we would like to discuss various dimensions and infrastructures in order to establish connectivity in the rural parts of India. We would also focus on the importance of maintenance of these wireless deployments and how it can be achieved.

\end{abstract}

\section{Keywords}
Rural Wi-fi, Mesh networks, Community co-design, Rural development, Wi-Fi enabled long distance networking (WiLD).

\section{INTRODUCTION}

Just a decade ago, our lives have been probably imagined as a “modern stone age”. Modern because the world was at a stage of evolution into something new. We should note that we did have computers jumbled with numerous wires. Wireless technology has become more prevalent today and continues to pervade the world of computer networks at a rapid pace. Now, since the wireless technology has emerged, it is possible to depict a clear picture of how the world has come so far ahead in ground breaking technologies.

Today, Network has assumed a new meaning, a new definition in people's lives. It has become an integral part of everyday life and has brought revolutionary changes in the way people think, communicate and connect with each other. A network can be anything if it's not for technology. For instance, in our daily lives, we need a network of roads to travel from one place to another. Similarly, to speak in terms of technology, a network of wires and antennas help us in transferring information from one place to another regardless of the geographical distance. With the help of computers and mobile devices, we can access data about anything in the world from the Internet. Landlines, broadband, and wireless connections where terrestrial connectivity or infrastructure is non-existent, all do their bit to bring the world to our doorsteps. Connecting people and communities is considered the hallmark of a developed information society. 

However, Internet penetration and connectivity in rural areas still pose a major challenge. There are myriad issues when we talk about providing connectivity to rural areas. To overcome this problem, a large number of efforts have already been made. The problem we are facing is about how to set up the infrastructure while minimizing the costs of deployments which is very important for a country like India. Free spectrum and Wi-Fi technology like Wimax are initiatives to bridge the divide. For example frequency bands in 2.4 GHz and 5.8 GHz are utilized to solve this problem. It is a free spectrum that can be used by anyone without taking a license or paying a fee to the government. The potential of this technology is realized in India and it is used to provide the connectivity to neglected rural areas. 

There is also an initiative launched by the Government of India which focuses on transforming the infrastructure of public services through the use of Information Technology. 

\begin{figure}[h]
\includegraphics[width=8cm,height=6cm]{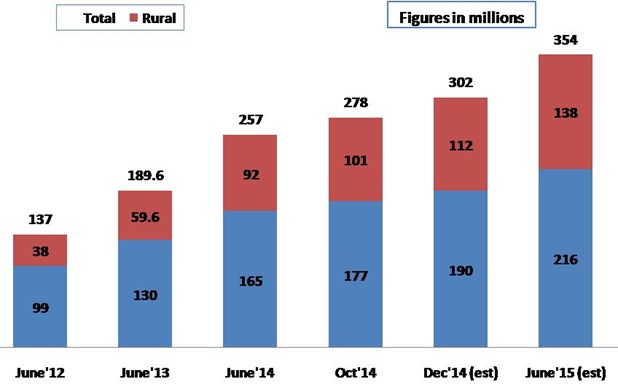}
 \caption{Internet Users Growth in India taken from IAMAI [14]}
\centering
\end{figure}

Internet Users in India crossed 300 million mark with significant growth of users in rural areas in December 2014 according to the report published by the Internet and Mobile Association of India (IAMAI) as shown in Fig.1 [14].

In this paper, we would like to focus on some of the methods for wireless deployments in rural India. In section 3 we will discuss the background motivation and in sections 4 to 6, we will discuss the challenges to setting up wireless networks in rural areas and the various implementation techniques to address them.

\section{Motivation}

India has the highest mobile penetration but teledensity in rural areas is still less than 40\%. Even though India has the third largest population of internet users, less than 12\% of its population is connected to the Internet. According to the Internet and Mobile Association of India (IAMAI), only 2\% of India's rural population has access to the web when almost 67\% of India’s 1.2 billion populous live in rural India [14]. Internet is not only the way to connect and communicate with friends, it also impacts the country's economy as it is responsible for providing a large number of jobs, helps small and mid-scale businesses, which in turn facilitates an increase in country's GDP. It also helps in solving the problems that society is facing today and acts as a bridge between technology and everything, be it arts, sciences, business etc. Thus, in a nutshell, it is the holistic approach combining people and technology bringing out the best from the world.

India is facing issues such as corruption, unemployment, and poverty and since the major populous live in the rural areas, it is very important for them to be connected to the internet which will in turn help in providing a lot of job opportunities and reducing unemployment. 

India is an agrarian society with 50\% of the workforce are farmers and accounts for around 13.7\% of the country's GDP [15]. The information required for the members of India's agrarian society such as crop prices, weather reports, new technologies are mostly not accessible to the farmers. Wireless deployments in the rural areas would help in bridging this gap.

The literacy rate of India is around 75\% which is majorly because of the fact that the majority of people living in villages and rural India either don't get access to proper education or are forced to start working at an early age. There is a scarcity of schools and colleges in rural areas and due to lack of primary education, people fail to understand the importance of Education. 
A survey named called the Annual Status of Education Report (ASER) [16], shows that even though the number of rural students attending schools is rising, but more than half of the students in fifth grade are unable to read the second grade textbook and are not able to solve simple mathematical problems.

With the help of Internet, people could be made more aware of the importance of education. They would also be able to access the education material through online universities and MOOC providers such as Courseera, NPTEL, Khan Academy [18,19,20].

\section{FACTORS AFFECTING RURAL WIRELESS DEPLOYMENTS}

There are various factors that need to be considered in order to provide low-cost, robust and rapidly deployable solutions in the Rural areas. Most of the users in rural India use cellular networks. The alternative to the cellular network is to make use of wireless mesh networking. The major research in providing wireless networks was focused on developing 802.11 based mesh networks along with other techniques to provide broadband connectivity to villages in developing regions and there are operational challenges faced during deployments of such networks.

\begin{figure*}
  \includegraphics[width=\textwidth,height=6cm]{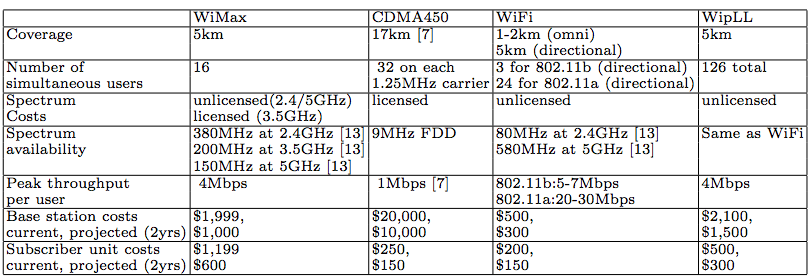}
  \caption{A comparison of capabilities and costs of various Access technologies done by Mishra et al. \cite{c4}}
\end{figure*}

\subsection{Power Issues}
The Power supply is one of the important factors that need to be considered while deploying rural wireless networks. As per records of Government of India, only 34,887 villages are yet to be electrified [9]. Moreover, the villages that are already electrified do not guarantee an uninterrupted power supply either. 
The unreliability of grid electricity in remote areas is one of the main challenges. Grid electricity is scarce and there are frequent multi-day power outages. Therefore, it is not feasible to rely on mains power and it becomes very important to have the battery powered network devices which can be used even when the grid supply is not present. 

These issues can be overcome through other alternatives such as powering the devices through Batteries. This was already done in one of the wireless deployment projects in South Africa where to overcome the power issues, the deployment team by Moreno et al[3] made use of 12V batteries for all mesh public phones in their project. They are mostly charged by solar panels through which they also addressed the questions raised by some environmentalists.

Another effort in this direction was made by AirJaldi to reduce the cost of networking is to switch its energy resources to solar. AirJaldi made the move because they realized that the cost of maintaining the power grid with backup diesel or batteries was very high.

Use of Solar power to solve the power issues may seem uneconomical but it actually turned out to be one of the main reasons for the success of Airjaldi while providing the internet connectivity with minimal down-time [2].

\subsection{Economic feasibility \& Cost of Network}

Economic feasibility and Cost of Network is an important factor in network deployment. Providing internet services in the rural areas given the low-income levels is a challenging problem. The existing connectivity technologies have high deployment costs making them unviable for rural areas. 

Due to high costs of setting up the wired networks, wireless networks are preferred in these areas. Wireless networks are relatively easy and quick to deploy, particularly in cases where we do not need new towers. Networks in unlicensed spectrum are preferred because they can be set up by grass-roots organizations as needed, avoiding dependence on a telecom carrier. This is particularly important for rural areas, which are less enticing to carriers due to the low income generation potential [6]. Connectivity solutions such as fiber-optic networks may not be feasible due to the cost factors. Also, it's not pragmatic to set-up high bandwidth links due to minimum bandwidth requirements by most of the applications. Therefore, it becomes an important requirement to make use of the technologies that minimizes the cost of deploying the wireless networks.

In the case study done on Akshaya Project in Kerala [4], following access technologies were considered WiFi, WiMax, CDMA450 and WipLL in combination with three forms of backhaul technologies: Fiber(PON), WiFi (with directional antennas) and VIP. In Fig.2 the researchers compared the characteristics and costs of these four technologies [4]. It was found that wireless network using WiFi for the backhaul, CDMA450 for the access network, and shared PCs for end user devices has the lowest deployment cost. If the expected spectrum licensing cost for CDMA450 is included, a network with lease exempt spectrum using WiFi for the backhaul and WiMax for access is the most economically attractive option [4].

There have been other cost effective initiatives such as Gyandoot, Drishtee, TARAhaat, Lokmitra that provided public community-based access to ICTs for educational, personal, social and economic development [2].

Airjaldi is also known for implementing the economically viable connectivity solutions for rural areas. It is one of the most advanced community-based networks in India. Airjaldi made use of two bandwidth ranges (2.4 GHz/5.8 GHz), known as Wi-Fi ranges, were delicensed by the Government of India [2].  

AirJaldi uses Ubuntu for capacitating it's networking operations, provisioning of client accounts, and remote server configuration and management. Since Ubuntu is a free, open-source operating system, the cost is negligible in comparison to Windows or Mac systems [2]. Airjaldi also makes use of solar energy resources to reduce their operational costs of maintaining a power grid.

\subsection{Quality of Service}

The deployment team in Subramanian et al [6] suggested the use of WiFi-based Long Distance (WiLD) links instead of Mesh networks as cost-efficient networking solutions for providing connectivity to regions with low user densities. There are many applications that use WiLD networks require QoS.

The researchers argued that Wild links are easy to deploy and experiment and they are cheap as they make use of the unlicensed spectrum. They differ significantly when compared with mesh networks in terms of external WiFi interference, multipath characteristics and routing protocol characteristics.

In Aravind Eye Hospital deployment, video-conferencing sessions with patients in rural areas is required where QOS is paramount importance. Although modifying routers to achieve QOS is possible in WiLD, a lot of QOS mechanisms do not carry over because of the constraints applied by WiLD networks. First, It is different from wired links and does not have the characteristic of fixed bandwidth value so any variation in the slot size along one link affects the one-way bandwidth on adjacent links. Second, providing end-to-end bandwidth and delay guarantees for flows requires scheduling mechanisms that can take into account the variable link bandwidths and link delays [6]. Flow isolation does not hold in WiLD since the resource allocation of competing flows can be affected by newly introduced flows. Also, because of low processing power and memory of WiLD nodes, some of the strict QOS mechanisms are ruled out. As per Subramanian et al.[6], the deployment team had already described the manipulations in primary link parameters to achieve the QOS.

\subsection{Local Cooperation \& Community Training Practices}

Involvement of the locals is a key factor in the success of wireless deployments. It is important to meet people and conduct interviews and surveys to understand and assess the possibility of deployment of the wireless network.

Building a network is a complex social process and not just a technical challenge. It is not possible to rely on the deployment team completely for the maintenance and administration of the wireless network. Active participation of the locals is a critical factor to the success of deployments. For a sustainable network establishment, conducting hands-on training is necessary. The main motive of conducting these training programmes is to make sure that the locals should manage to fix the issues themselves when the need arises. They should also be able to build the competencies with experience such that communities should be able to build and govern their own communications infrastructure. The Open Technology institute in collaboration with Airjaldi conducted workshops in India and Nepal to train communities to build and manage the wireless networks [8].

\subsection{Safety} 
Once installed, devices can also suffer from harsh environments that may destroy them. Especially in areas with harsh terrain and weather conditions, the safety of network nodes is very essential. Dharamshala is located in the foothills of Himalayas. In order to combat the harsh environment mainly due to wind, and rain, additional effort has been made to buy most of the auxiliary materials that are rust-proof. Moreover, the wind has to be handled to limit the movement of antennas so that they are not misaligned. 
Additional measures are required in securing the nodes to prevent people from damaging or stealing them. The deployment team in Gabale et al.[1] was concerned about the safety of gateway node because of the gateway crash incident. To tackle this they are planning to  lock the gateway node in a small compartment to protect it from such damages.

\subsection{Cultural Issues} 
According to the author in Moreno et al.[3], a wireless network deployment project in South Africa, both men and women were trained for the project. It was noticed that there was a drop in the participation of women in the later stages even when they appeared enthusiastic and engaged in the first few sessions. Since women had other responsibilities like childcare and domestic work, sustained participation was difficult. However, it was not the same case with men who made through to the later stages.

\section{Maintenance of Rural Wifi Networks}
There is a lot of work and research that has been done on wireless networking technologies [5] in rural areas but one factor which is majorly overlooked is the ability of local staff to fix errors which result into excessive outages. There are various reasons why maintenance in rural areas is difficult with the primary reason being the limited technical knowledge of local staff about wireless networks. The other reasons that make maintenance difficult are unreliable power supply that leads into a lot of hardware failures and also the location of wireless nodes as they are located in remote areas, so it makes frequent visits to these locations expensive and impractical. It is also possible that a single broken link makes the whole network unreachable even though the nodes themselves are functional and running which makes it very difficult for the administrator to troubleshoot and resolve the issue.

When the community is not technically competent enough to resolve the issues by themselves and it's not feasible for the deployment team to visit the site every time when there is an issue, We require a solution that can help the deployment team to manage the network locally with the minimal help of locals. Analyzing the fault situations arose from Aravind Eye Hospital telemedicine project, the research team identified the key factors in order to implement an end-to-end diagnostic system for rural wireless networks. 

The researchers presented some of the requirements of monitoring the networking efficiently.

\subsection{Monitoring Status}
It is essential to monitor the status of nodes in the network regularly such that in the case of a failure in any node, fault localization is easy without requiring the admins to query all the nodes manually to find the point of failure in the network. 

The admins need an infrastructure that continuously probes all the nodes and relays (e.g. ping tests) and presents the current reachability of the network in a graphical form at a central place. Finally, when nodes or relays go down, admins should be notified that part of the network is down for possibly proactive action [5]. 

\subsection{Predicting Behaviour}
Due to unreliable grid power, batteries are used in most of the relay points in the deployment network in [5]. After a certain amount of time, these batteries discharge completely which causes the network to go down. Therefore, a framework is required to predict the immediate remaining uptime of the network and schedule the opening or closing of the video conferencing [5] between rural patients and doctors at the main hospital for remote eye-care consultations.
It is also essential for efficient monitoring to predict the expected lifetime of batteries by keeping a count of deep discharges or by tracking how the rate of discharge changes over time [5].

\subsection{Comparing Expected Behaviour}
It is possible that the network is down but it could be verified that the routers are working properly which shows that there might be a problem with the antenna. It could be due to antenna misalignment or that some antenna cable or connector is disconnected. The research team in [5] stated that it is possible to identify some of the causes by comparing the measured signal strength before and after the problem. The antenna alignment status can be determined by observing the signal strength. Therefore we can deduce by citing the antenna example that, for efficient monitoring, the system should have the ability to compare current behavior with generally expected behavior and derive conclusions from it.
\\
\\
The researchers also stated the independent hardware modules that facilitate diagnosis by fulfilling monitoring requirements.

\subsection{Back Channels}

An independent back channel is required to diagnose the network when the primary wireless link goes down. It is hard to debug these networks because the only links to remote nodes are the wireless links themselves. As a result, we need to build alternate mechanisms to reach remote nodes or query systems inaccessible by primary links [5].
The remote node could become inaccessible due to network misconfigurations or the problem could be more serious than routing misconfiguration. Therefore, a system is required to access the remote nodes independently from the primary wireless link.
This can be achieved using a mechanism such as Local Link Addressing where each link gets a local automatic IP address from a pre-assigned subnet that would work even
when the system-wide routing does not work [5].
Another alternative is using SMS on a mobile network which consists of an SMS channel over the cellphone network which can be queried in the case of failure of primary wireless link [5].

\subsection{Separate Hardware Control}
An independent hardware-based module that reboots the system when it does not receive periodic heartbeats [5]. Due to poor power or crashing of the wireless driver, a link can go down which requires rebooting of the system. It is not feasible to go to the remote node just for rebooting. Researchers in [5] suggested independent hardware components that are used to reboot the router board when an anomalous state is detected. A Hardware watchdog is a form of independent control on the board itself which reboots the board if the OS fails to poke a particular register at a regular interval. Local Power Controller is another option that monitors the board using a serial port and reboots it whenever it does not receive a response from the board on the serial port [5].

\subsection{Software Services}
The research team of Surana et al.[5] came up with different software mechanisms to support safe upgrades and restart failed services. They made use of software watchdog service would periodically use heartbeats to a set of services such as the routing daemon, DTN daemon (for logs) and restart them on detecting a crash. They achieved safe upgrades by making sure that on an upgrade of the router OS, the service is configured to look for parameters which ensures that the upgrade does not violate any required properties.

\section{Case Studies}
Various initiatives have been taken by the organizations to facilitate the process of wireless deployments in Rural India such as Wireless for Communities (W4C), an initiative of Digital Empowerment Foundation (DEF), AirJaldi, the brand name used by Rural Broadband Pvt. Ltd., Akshaya Project.

\subsection{Wireless for Communities}
Wireless for Communities (W4C) is an initiative taken by Digital Empowerment Foundation (DEF), the Internet Society (ISOC) and Ford Foundation to connect rural and remote locations of India. They target the areas where the mainstream Internet service providers don't provide the connectivity because of not so much commercial benefits. W4C makes use line-of-sight wireless technology and low-cost Wi-Fi equipment which utilizes the unlicensed 2.4 GHz and 5.8 GHz spectrum bands to create community-owned and operated wireless networks. The aim is to provide training to the community members about wireless technologies so that they gain competency to run and maintain the community networks and they are also trained to pass on their knowledge to other members to be able to build a sustainable system.

The first pilot project was started in October 2010 in Chanderi district which is famous for its traditional silk weaving art. The success of this project motivated the other stakeholders to replicate this model in the second phase which focused on the wireless deployments in western desert state of India called Rajasthan for tribal communities. Phase 2 program was mainly implemented in Baran, Tilonia in Rajasthan and Tura in Meghalaya. In Phase 3, the W4C Program was implemented in Giridih (Jharkhand), Mandla (Madhya Pradesh) and Naogang (North Tripura). These are amongst the remotest and most backward of regions in India [7]. 

The W4C project has been a success and they are able to spread to eight locations and benefited thousands of users – all of them in remote areas. The efforts have attracted the attention of other stakeholders, including policy advocates, government and private players to adopt wireless technology as an alternative solution towards connectivity and access.

\subsection{AirJaldi}
AirJaldi, an initiative by Rural Broadband Pvt. Ltd., provides high-quality broadband connectivity to rural areas at reasonable rates. AirJaldi currently serves 70,000 Indians who previously had little to no internet connectivity and aims to reach 25\% of India [12]. Their main goal was to design and build customized gear that can take the wear and tear of the rough rural conditions. The system uses available solar energy to power the network backbones and end equipment which is not dependent on the availability of power in a rural area. Airjaldi also used the local community to set up the infrastructure by providing subsidized connectivity to these recipients and in turn sharing/utilizing their existing assets such as space, power, and local resources in return for providing affordable internet.
AirJaldi has helped the communities living in rural hilly areas and the Tibetans in exile by building and deploying the wireless networks. 

In the beginning, AirJaldi started providing broadband internet connectivity to schools, local communities and monasteries in the Dharamshala area and also trained the community people with technical knowledge and basic computer and wireless networking. At present it provides connectivity to four rural states in India what seemed impossible initially. The initiative has empowered these communities by providing economical access to the internet [2].

\subsection{Akshaya}
Akshaya Project is the joint initiative by Tulip IT and the Government of Kerala, India, to provide rural wireless connectivity to Malappuram district in Kerala. Malappuram is now said to be India's First E-literate District. It is claimed to reach over 600,000 households, representing more than 3.6 million people [4]. The project aimed to provide connectivity to the group of people rather than individual users. The network is maintained by local entrepreneurs who in turn are aided by the Government as the loans are sanctioned by the Government to support the initiative. The Akshaya network make use of VINE (Versatile Intelligent Network Environment) technology for wireless backhaul links which is incorporated in VIP radios powered by Wi-LAN, Canada. The VIP node can either serve as a basestation or as a repeater. Each radio has 1W transmission power and provides 11Mbps in the 2.4GHz band. For connecting the backhaul network with the Akshaya Centers, (point to multi-point links) Tulip used WipLL technology from Marconi and Airspan [11].

\subsection{Facebook}
Facebook too came up with a mission of providing free Internet access in rural India.  The idea was to provide free access to health, education, local and national news through an Internet connection which could potentially improve the quality of lives. Facebook’s proposed Free Basics plan that allows customers to access the social network and other services such as education, health care, and employment listings from their phones without a data plan. The activists refused to accept the plan saying this could threaten the principles of net neutrality. However, the plan failed as the People of India voted against it resulting into TRAI (country's regulating authority) banning it. They believed that Internet.org [21] is going to kill the open nature of internet since some of the websites will be favored over the others. There were protests on the streets and online campaigns like “Save the internet” were launched. The ruling by the Telecom Regulatory Authority of India (TRAI) now forbids all zero-rating plans, that means anyone offering customers free access to only a limited set of services of sites is banned [13].

\section{CONCLUSION}
In this paper, we have reflected on the efforts made to deploy and manage the wireless networks in India. We also discussed the challenges faced by various deployment teams in laying down the infrastructure. To this end, various wireless options are examined and explored. We also discussed their suitability based on the needs. It is observed that community support is paramount in the success of these deployments. The paper also presents the case for economically viable networks in rural developing regions. The experiences of various organizations documented here on network planning and network sustenance would be useful for future systems targeted for rural deployment.

\addtolength{\textheight}{-12cm}   



\end{document}